\documentclass[twocolumn,preprintnumbers,secnumarabic,amsmath,amssymb,superscriptaddress,nofootinbib]{revtex4}
\usepackage{graphicx}% Include figure files
\usepackage{dcolumn}% Align table columns on decimal point
\usepackage{bm}% bold math
\usepackage[utf8]{inputenc}
\usepackage{hyperref}
\newcommand\numberthis{\addtocounter{equation}{1}\tag{\theequation}}

\newcommand{\nn}{\nonumber}

\newcommand{\beq}{\begin{equation}}
\newcommand{\eeq}{\end{equation}}

\newcommand{\bea}{\begin{eqnarray}}
\newcommand{\eea}{\end{eqnarray}}

\newcommand{\ii}{\text{i}}

\def\ZZ{\mathbb{Z}}

\begin{document}

\title{$D3$-Brane Model Building and the Supertrace Rule}

\smallskip
\author{Iosif Bena, Mariana Gra\~na, Stanislav Kuperstein, Praxitelis Ntokos}

\affiliation{{Institut de Physique Th\'eorique, CEA
Saclay, CNRS UMR 3681, F-91191 Gif-sur-Yvette, France
}}

\author{Michela Petrini}

\affiliation{{LPTHE, UPMC Paris 06, UMR 7589, 75005 Paris, France\\  \textsf{iosif.bena, mariana.grana, praxitelis.ntokos@cea.fr, stanislav.kuperstein@gmail.com, michela.petrini@lpthe.jussieu.fr}
}}

\preprint{IPhT-T15/183}

\begin{abstract}

A common way to obtain standard-model-like Lagrangians in string theory is to place $D3$-branes inside flux compactifications.
The bosonic and fermionic masses and couplings of the resulting gauge theory are determined by the ten-dimensional metric and the fluxes, respectively, and the breaking of supersymmetry is soft. However, not any soft-supersymmetry-breaking Lagrangian can be obtained this way since the string theory equations of motion impose certain relations between the soft couplings. We show that for $D3$-branes in background fluxes, these relations imply that the sums of the squares of the boson and of the fermion masses are equal and that, furthermore, one- and two-loop quantum corrections do not spoil this equality. This makes the use of $D3$-branes for constructing computationally controllable models for physics beyond the standard model problematic.

\end{abstract}

\maketitle

$D$-branes provide a very nice mechanism for embedding supersymmetric gauge theories in type II string theory.
There is an extensive literature on using branes extended along a (3+1)-dimensional space and wrapping some cycles (or a point) in a six-dimensional  (internal)  manifold 
to construct four-dimensional effective theories that have a field content similar to that of the standard model (for reviews, see Refs. \cite{braneSM1,braneSM2,braneSM3}). In these constructions, the low-energy excitations on the branes give the gauge-theory sector, 
with  masses and couplings related to the low-energy closed string modes of the internal manifold. 

Phenomenologically relevant models arise when nontrivial background fluxes on the internal space are turned on. Whenever these fluxes break supersymmetry  in the bulk, this is communicated to the gauge sector through the bulk fields (via, for example, gravity-mediation supersymmetry-breaking scenarios), generating soft terms for the matter fields. 

The main advantage of soft-supersymmetry breaking compared to spontaneous breaking is that the former can avoid the supertrace sum rule
 \beq \label{supertrace}
 %{\rm Str}\,  M^2=0 \quad \textrm{or} \quad 
 \sum_\textrm{bosons} m_b^2 =  \sum_\textrm{fermions} \!\!\! m_f^2 \, ,
\eeq
and hence avoid the existence of supersymmetric particles much lighter than the top quark, which is essentially ruled out by recent LHC results. 

The simplest low-energy theories can be obtained using $D3$-branes transverse to the six-dimensional manifold - these are $U(N)$ gauge theories 
 whose field content and symmetries  are determined by the geometry of the internal space. To obtain theories that are more relevant phenomenologically, one usually places the $D3$-branes 
at  singularities in the internal space, which breaks the $U(N)$ gauge symmetry  into standard-model- or grand-unified-theory-like gauge groups. 
As already mentioned, fluxes on the internal manifold induce  soft-supersymmetry-breaking terms in the gauge theory  \cite{Grana:2003ek,Camara:2003ku} and one may therefore hope to use these branes to construct realistic models of physics beyond the standard model  (BSM). 

The purpose of this Letter is to show that, even if the breaking of supersymmetry on the $D3$-branes is soft, the soft terms still obey \eqref{supertrace}, not only at tree level, but also (at least) at one  and two loops.
Hence, the tree-level zero-supertrace condition appears to be a universal feature of any $D3$-brane in equilibrium in a flux compactification whose metric, dilaton and fluxes obey the equations of motions of supergravity.  By explicit calculations we checked that the supertrace also vanishes at one and two loops when the $D3$-branes are at a generic minimum, and at one loop when the D3-brane is on top of $\ZZ_2$ and $\ZZ_3$ 
orbifold singularities. This appears to be a feature of other $\ZZ_N$ singularities as well. Hence, our result indicates that any field theory built using  
such $D3$-branes will have this feature and hence will not be a feasible candidate for describing BSM physics.  The only way to avoid this problem would be to rely exclusively on nonperturbative corrections, though, as we will explain in the section about quantum corrections,  it is not clear whether these corrections can do the job.
\vspace*{-.3cm}
\section{Softly broken $\mathcal{N}=1$ theories}
\vspace*{-.3cm}

We are interested in ${\mathcal N}=1$ theories that descend from ${\mathcal N}=4$ super Yang-Mills (SYM) theory, which can be found on the world volume of $D3$-branes extended along the spacetime directions and sitting at a point in some six-dimensional compactification space. These theories have  three chiral multiplets $\Phi^i$, $i=1,2,3$, transforming in the adjoint representation of the gauge group $U(N)$ (for $N$-branes) and a superpotential 
\beq \label{W}
W=\frac12 m_{ij} \Phi^i \Phi^j + \frac16 g \epsilon_{ijk} \Phi^i \Phi^j \Phi^k \ ,
\eeq
where, to simplify the notation, we have omitted the trace over the color indices. The last term is the superpotential of the  original  ${\mathcal N}=4$ SYM, and the first one corresponds to a generic mass term  that, as we will see, is generated by the fluxes on the six-dimensional space. Supergravity fluxes can also induce soft-supersymmetry-breaking terms.  Generically, the  Lagrangian containing both supersymmetric and  soft-supersymmetry-(SUSY)-breaking terms has the form (up to cubic terms) 
\vspace*{-.3cm}
\begin{align*}
 & {\cal L}_\text{SUSY}+{\cal L}_\text{soft}  =  - (m m^{\dagger})_i{}^{j}     \phi^i \bar \phi_j - \left(\frac{1}{2} m_{ij} \psi^i \psi^j + \textrm{H.c.} \right) \\
& - (m^2_{\text{soft}})_i{}^j \phi^i \bar \phi_j - \left( \frac{1}{2}  b_{ij} \phi^i \phi^j +   \hat m_{i}  \psi^i \lambda  +  \frac{1}{2} \tilde m \lambda \lambda +\textrm{H.c.}  \right)  \\
& - \Big( \frac{1}{2} m_{il} \epsilon^{ljk} \phi^i \bar \phi_j \bar \phi_k  \numberthis \label{LsusyLsoft} \\
& \qquad + \frac12 c_{ij}^k \phi^i \phi^j \bar \phi_k + \frac16 a_{ijk} \phi^i \phi^j \phi^k  + \textrm{H.c.} \Big) . 
\end{align*}
\vspace*{-.35cm}

\noindent Here, $\phi^i$ and $\psi^i$ are the bosonic and fermionic components of the chiral field $\Phi^i$ and $\lambda$ is the gaugino. The first and third lines contain the supersymmetric terms coming from the superpotential \eqref{W}, while the second and fourth are soft-supersymmetry-breaking terms:
bosonic masses, scalar bilinear terms, quadratic couplings between  the chiral fermions and the gaugino\footnote{These are not usually considered in the literature as there are no chiral fermions transforming in the adjoint representation. We will show that  the fluxes giving rise to these terms are not allowed in the most typical situations.}, gaugino mass,
trilinear $c$, and A terms\footnote{The $c$ term is not usually considered because it can lead to quadratic divergencies if there are gauge singlets. For $D3$-branes in fluxes, it arises from the same fluxes as $\hat m_i$.}. 

More interesting models for phenomenology are obtained using singular six-dimensional manifolds  
and putting branes at the singularities.  A simple class of such models comes from $\ZZ_p$  orbifolds of six-dimensional flat space for which the gauge symmetry is enhanced to $U(p N)$, where $p$ is the number of images of a single brane under the $\ZZ_p$ symmetry that gives the singularity, and then broken to subgroups of $U(p N)$ by splitting the branes and  image branes in  different stacks \cite{Douglas:1996sw}.  For the simplest example of a $\ZZ_2$ singularity with $N$-branes one obtains in the end a theory with a $U(N) \times U(N)$ gauge group, while  the adjoint matter of the original theory  splits into two bifundamentals of the two different  gauge groups plus two adjoint chiral fields, each charged under one of the gauge groups.  

Knowing the action of the symmetry group at the singular point  allows one to obtain the Lagrangian of the orbifolded theory  from that of the ``original" SUSY  
${\mathcal N}=1$ theory \eqref{W}.  The  structure of the softly broken theory is the same  as in  \eqref{LsusyLsoft}, where now the  matter fields
are in the bifundamentals of the different gauge groups.   Generically, the orbifold symmetry constrains some of the couplings in 
 \eqref{W} or  \eqref{LsusyLsoft} to be equal for the different gauge groups, or to be zero if they do not respect the symmetry. 

\vspace*{-.3cm}
\section{Susy and soft terms for D3-branes in fluxes}
\label{sec:The-Mass-Deformation}
\vspace*{-.3cm}

In the gauge theories that live on the world volume of $D3$-branes in flux backgrounds, both the supersymmetric masses $m_{ij}$ and the soft-supersymmetry-breaking terms arise from the supergravity fields. These are the ten-dimensional metric $g_{10}$, the dilaton $\phi$, as well as the gauge fields: a pair of two-form gauge fields $B_2$ and $C_2$, and a four-form $C_4$.  
It is convenient to combine the  field strengths of  $B_2$ and $C_2$ into a complex three-form 
\beq \label{G}
G_3=F_3 - i \, e^{-\phi} H_3  = d C_2 -  i \, e^{-\phi}  d B_2\ ,
\eeq 
which will play a crucial role in the gauge-theory Lagrangian.  A general string theory compactification with fluxes has a warped metric of the form
\beq
g_{10}(x,y) = \begin{pmatrix} e^{2 \alpha (y)} g_{{4}}(x) &  0 \\ 0 &  g_6(y) \end{pmatrix} \ 
\eeq 
where $x$ and $y$ are the four (``external") and six (``internal") coordinates, $\alpha$ is the warp factor, and $g_{{4}}$ is the four-dimensional Minkowski metric.

The supersymmetric masses $m_{ij}$ and the soft fermionic masses $\hat m$ and $\tilde m$ are generated by the bulk supergravity fields, and the precise relation between them is most easily obtained by computing the fermionic $D$-brane action \cite{Grana:2002tu}.  It is important to note that the
${\cal N}=1$ gauge theory \eqref{W} descends from ${\cal N}=4$ SYM and hence it has a memory of the original $SU(4)$ $R$ symmetry of the ${\cal N}=4$ theory.
Specifically,  the  three fermions $\psi^i$ in the chiral multiplets can  be combined  with the gaugino $\lambda$ to reconstruct the ${\cal N}=4$
fermions in the ${\bf 4}$  of $SU(4)$.  Then the  fermionic masses can similarly be combined into a $4 \times 4$ mass matrix
\beq \label{MSU3}
 M_{IJ} = \begin{pmatrix} m_{ij} &  \hat m_i \\ \hat m_{i}^T &  \tilde m \end{pmatrix} \ 
 \eeq 
with $I=1,\dots, 4$. This matrix transforms in the ${\bf 10}$ of $SU(4)\cong SO(6)/\ZZ_2$, and it can equivalently be encoded in an imaginary anti-self-dual [IASD: $(*T)_{ABC}\equiv \frac{1}{ 3!} \epsilon_{A B C}{}^{D E F} T_{D E F} = - i T^{A B C}$]  three-form on the six-dimensional space, $T_{ABC}$ ($A,B,C=1,\dots,6$) which also has ten independent components  
%\footnote{\label{foot:ISD}In our conventions  $(*T)_{ABC}\equiv \frac{1}{ 3!} \epsilon_{A B C}{}^{D E F} T_{D E F} = - i T^{A B C}$.} 
and transforms in the ${\bf 10}$ of $SO(6)$. The map between the mass matrix and this three-form is given by
\begin{equation} \label{TM}
\begin{split}
T_{A B C} &= -\frac{1}{2 \sqrt{2}} \textrm{Tr} \left( M \eta^A {\eta^B}^{\dagger} \eta^C \right) \, ,  \\ 
M_{IJ} & = \frac{1}{12 \sqrt{2}} T_{ABC} ({\eta^A}^{\dagger}  {\eta^B} {\eta^C}^{\dagger} )_{IJ} \, ,
\end{split}
\end{equation}
where the six matrices $\eta^A$ that intertwine between $SU(4)$ and $SO(6)$ are usually called 't Hooft symbols, or generalized Weyl matrices 
(their  explicit expression  is  given in the Appendix),
and the numerical coefficients are chosen to match the conventions of Ref. \cite{Polchinski:2000uf}. The splitting of the matrix $M$ into its ${\cal N}=1$ components (\ref{MSU3}) corresponds to selecting one supersymmetry among the four of ${\cal N}=4$ SYM theory and is equivalent to choosing a set of complex coordinates on the six-dimensional internal space.
The IASD three-form $T$ splits into components with different numbers of holomorphic and antiholomorphic indices. The fundamental $SU(4)$ index $I$ splits under $SU(4)_R \to SU(3) \times U(1)_R$ into  $I=(i,4)$ and the  fundamental $SO(6)$ index $A$ splits into  $ A=(i, \bar \imath)$. We thus find
\vspace*{-.2cm}
 \bea \label{TandM}
m_{il} &=&   \tfrac12 T_{i\bar \jmath \bar k} \epsilon^{\bar \jmath \bar k}{}_l    \nn \\  
\hat m_{i} &=&- \tfrac{i}{2} T_{i j \bar k} J^{j \bar k}  \\  
\tilde m &=& \tfrac16 T_{i j  k} \epsilon^{ijk}   \nn 
\eea
where $J_{i\bar \jmath}$ is the symplectic structure associated with the choice of the $SU(3)$ subgroup (in our conventions  $J_{1 \bar 1}=J_{2 \bar 2}=J_{3 \bar 3}=i$).

For $D3$-branes in flux backgrounds, the tensor $T$ is the IASD piece of the complex three-form flux $G_3$ introduced in Eq.  (\ref{G}) of Ref. \cite{Grana:2002tu}
\beq
\label{T3}
T_3=e^{4 \alpha}( \star_6 G_3 - \ii G_3)\, .
\eeq  
Here, we have used the notation of Ref. \cite{Polchinski:2000uf} and  the Hodge star $\star_6$ on the six-dimensional space is defined below Eq. (\ref{MSU3}). $D3$-branes in Calabi-Yau manifolds 
have a moduli space corresponding to the fact that the brane can sit at any point inside 
the Calabi-Yau (CY) compactification.  The same is true if one adds to the background an imaginary self-dual  flux $G_3$.  
However, introducing  an IASD component  generically uplifts this moduli space and, as a result, the branes only have a finite number of minima. 
The equations of motion imply that the tensor $T_3$  is  position independent, and therefore the masses are the same, regardless of where the branes sit.

From Eqs. (\ref{TandM}) and (\ref{T3}), we see that  
the (1,2) component of the IASD fluxes gives rise to $m_{il}$, which can be included in a supersymmetric Lagrangian, while the (3,0) component gives a gaugino mass $\tilde m$ that breaks supersymmetry softly on the brane. The flux terms that would give rise to $\hat m_i$ cannot arise in fluxed Calabi-Yau compactifications \cite{Grana:2000jj}, but they can appear when the branes are in more general backgrounds.

The soft SUSY-breaking trilinear terms $c$ and $a$ in Eq. (\ref{LsusyLsoft}) were computed in Ref. \cite{Myers:1999ps} using the bosonic non-Abelian $D$-brane action and are again entirely determined by $T_3$:
\beq \label{ca}
c_{ij}^k= T_{ij\bar k}= \delta^k_{[i} \hat m_{j]} \ , \quad a_{ijk}=T_{ijk}=\tilde m \epsilon_{ijk} \ .
\eeq
Note that, in a general theory with soft-supersymmetry-breaking, there is no relation between the boson trilinear couplings and the fermion masses, but in the theories that live on the world volume of $D3$-branes, these are always linked: the fermion mass matrix completely determines the boson trilinear couplings. This relation is a crucial ingredient of the calculation that establishes our result.
 
The scalar masses and the $b$ terms in Eq. \eqref{LsusyLsoft} are trickier to determine.  In the $D3$-brane world-volume action they are obtained from the potential 
felt by the $D3$-brane, $V = e^{4 \alpha}-\mathcal{C}$, where $e^{4 \alpha}$ is the warp factor and $\mathcal{C}$ is the four-form potential $C_4$ along the spacetime directions. By Taylor expanding $V$  around one of its minima,  $y^0$,  
\bea
\begin{split} \label{Taylor}
e^{4 \alpha}-\mathcal{C}= &(e^{4 \alpha}-\mathcal{C})|_{y^0} \\
& + \frac12 \partial^2_{AB} (e^{4 \alpha}-\mathcal{C})|_{y^0} (y-y^0)^A (y-y^0)^B + \dots, 
\end{split}
\eea
and identifying the distance to the brane with the scalar fields on the brane $(y-y^0)^A\sim \phi^A$, one can calculate all of the $(6 \times 7)/2=21$ entries of the matrix $\partial^2_{AB} (e^{4  \alpha}-\mathcal{C})|_{y^0}$, which give the 21 boson masses \cite{Grana:2003ek}. When choosing complex coordinates the boson masses split under $SU(4)_R \to SU(3) \times U(1)_R$ as 
%\begin{equation}
${\bf 20} + {\bf 1}={\bf 8}+{\bf 1}+{\bf 6}+ {\bf \bar 6}$,
%\end{equation}
corresponding, respectively, to the traceless part of $m^2_{i \bar \jmath}$ and its trace, as well as to $b_{ij}$ and its complex conjugate:
\beq \label{mlaplacian}
(m^2_{\text{SUSY}}+m^2_{\text{soft}})_{i\bar \jmath}=\partial^2_{i \bar \jmath} (e^{4 \alpha}-\mathcal{C})|_{y^0} \ , \ b_{ij}=\partial^2_{i j} (e^{4 \alpha}-\mathcal{C})|_{y^0} \ .
\eeq

Unlike the trilinear terms, which are completely determined by the background three-form fluxes, only one of the 21 boson mass components can be related to these fluxes, via the equations of motion. When the branes are at a minimum of the potential $V$,
a particular combination of the bulk equations of motion [see Eq. (2.30) of Ref. \cite{Giddings:2001yu}] allows one to fix  the trace of the boson mass matrix in terms of the three-form $T_3$ that determines the fermion masses:
\begin{equation}
\label{eq:BoxPhi-}
\nabla^2 ( e^{4 \alpha} -\mathcal{C})= 
\tfrac16 e^{4 \alpha+\phi} \left\vert \star_6 G_3-i G_3 \right\vert^2 = \tfrac16 e^{-4 \alpha+\phi}\ \left\vert T_3 \right\vert^2   \, 
\end{equation} 
 where $|T_3|^2=\frac16 T_{ABC}\bar T^{ABC}$. The other 20 components of the boson masses are not fixed by $T_3$, but are determined by the features of the geometry near the location of the brane. Thus, they are both model and location dependent. 
 
Using Eqs. (\ref{mlaplacian}), (\ref{eq:BoxPhi-}), and (\ref{TandM}), we finally arrive at the main formula of this Letter:
\bea
\text{Tr} (m^2_{\text{SUSY}} + m^2_{\text{soft}}) &=& \text{Tr} (m m^{\dagger}) + 2 \, \hat m_I \bar{\hat m}^I + \tilde m^2  \nonumber  \\
& =& \text{Tr} (M M^{\dagger})\ ,
\eea
where $M$ is defined in Eq. \eqref{MSU3}. In other words, we find
\beq \label{equaltraces}
\text{Tr} [\text{boson\,masses}^2]=\text{Tr} [\text{fermion\,masses}^2] \ .\\
\eeq

Hence, the actions of all $D3$-branes extended along the spacetime directions and sitting at equilibrium inside the compact manifold obey, at tree level, the zero-supertrace sum rule (\ref{supertrace}). In light of the recent LHC results, this makes problematic the use of these $D3$-branes in constructing minimally extended BSM models\footnote{The trace of the fermion masses includes that of the gauginos. However, having the latter heavy enough to overcome the problem of the supertrace rule brings in the little hierarchy problem and is therefore not a way out.}. We explain in the next section that this feature persists when one- and two-loop quantum corrections are taken into account.

We close this section with a comment. All of the parameters in the Lagrangian (\ref{LsusyLsoft}) can alternatively be computed purely within ${\cal N}=1$ supergravity with chiral fields including a hidden sector (moduli) on top of the observable sector (brane fields). After breaking supersymmetry spontaneously in the hidden sector via $F$ terms (which can be done by turning on three-form fluxes), integrating out the moduli fields and taking the limit of infinite Planck mass while keeping the gravitino mass finite, one obtains a softly broken  ${\cal N}=1$ gauge theory for the visible sector. The parameters of the latter are given in terms of the $F$ terms, the superpotential, and the K\"ahler potential of the original ${\cal N}=1$ supergravity theory\footnote{For $D3$-branes in CY compactifications, the K\"ahler potential is of sequestered form if the complex structure moduli are integrated out \cite{Grana:2003ek}, as done in Kachru-Kallosh-Linde-Trivedi models \cite{KKLT} or in large volume scenarios \cite{LSV}.}  \cite{KL,BIM}. Comparing these with those obtained from the $D3$-brane action, one finds \cite{Grana:2003ek} that they all agree, except for the boson masses. Furthermore, it is only for nonscale supersymmetry breaking and zero supersymmetric masses that the supertrace obtained by the supergravity calculation is zero;  generically, it is not. It would be interesting to understand why the supergravity calculation fails to reproduce this feature of the $D3$-brane action.  

\vspace*{-.3cm}
\section{Quantum corrections}
\vspace*{-.3cm}

The one-loop beta functions for all of the coupling constants including the ``nonstandard soft-supersymmetry breaking" terms $\hat m$ and $c$ in Eq. (\ref{LsusyLsoft})  were computed in Ref. \cite{Jack:1999ud}. By using the relation between the soft trilinear terms and the fermion masses (\ref{ca}), we find that all of the one-loop beta functions, except the ones for the boson masses, vanish exactly. The one-loop beta function for the trace of the boson mass matrix also vanishes if and only if Eq. (\ref{equaltraces}) holds, which is precisely what happens for the $D3$-brane
world-volume theories. We have checked this for branes at a regular point of the internal manifold, and also for branes at $\ZZ_2$ and $\ZZ_3$ singularities.

The two-loop beta functions were computed in Refs. \cite{MV} and \cite{JJ2}. We find  that for $D3$-branes at nonsingular points in the internal manifold, all of these beta functions again vanish when the supertrace of the square of the masses vanish (there might be additional regularization scheme-dependent conditions; for example, in Ref. \cite{JJ2} the mass of the fictitious ``$\epsilon$ scalar" should be set to zero).

It is very likely that all beta functions vanish perturbatively at all loops. Indeed, the fermionic masses (\ref{TandM}) are given by a constant (position independent) tensor, and therefore we do not expect them to run with the energy scale (corresponding to the radial distance away from the branes). Furthermore, since the trace of the bosonic masses is equal to the trace of the fermionic ones classically and at one and two loops, and the latter do not run, we expect this equality to hold at all loops. When the branes are placed in an $SO(3)\times SO(3)$ invariant background that  has (1,2) but no (3,0) components, this expectation can also be confirmed by explicit calculations \cite{Bena:2016fqp}: the theory on their world volume is simply $\mathcal{N}=4$ broken to $\mathcal{N}=1$ by the introduction of supersymmetric chiral multiplet masses, and it is broken to $\mathcal{N}=0$ only by a certain traceless bosonic bilinear. Using some clever superspace tricks, this theory was shown in Ref. \cite{Parkes:1983ib} to have vanishing beta functions at all loops. 

The expression for the soft parameters in terms of the fluxes is also expected to receive $\alpha'$ corrections coming from higher derivative terms in the ten-dimensional bulk and brane actions. These terms induce corrections to the K\"ahler potential of the four-dimensional  ${\cal N}=1$ supergravity theory that generically break the no-scale structure \cite{BBHL}, and induce corrections to the soft masses \cite{CQ} (and thus to their trace). In the so-called ultralocal limit, where the coupling between the matter and the moduli in the K\"ahler potential has a particular form, the breaking of no-scale structure is not seen in the visible sector, and the soft terms do not get corrected. From our arguments before, it is very likely that the full K\"ahler potential for $D3$-branes in Calabi-Yau manifolds falls into this category, and  our result holds even when taking $\alpha'$ corrections into account. On the other hand, one might expect that nonperturbative corrections to the superpotential, which are usually invoked in string phenomenology scenarios, modify this result. Unfortunately, there is no way of analyzing this at the ten-dimensional level, as such corrections are modeled in the four-dimensional field theory only. One could thus try to compute the soft terms, including these types of corrections using ${\cal N}=1$ supergravity calculations, as discussed in the previous section, and check to see whether the zero supertrace result still holds. However, it is hard to extract meaningful conclusions from such calculations: first, because these calculations fail already at tree level to reproduce the trace of soft masses found from the ten-dimensional equations of motion and, second, because to do these calculations correctly one would need to include the full dependence of the nonperturbative corrections on the moduli (particularly the unknown dependence on complex structure moduli). 

It is worth stressing that our analysis also holds for $D3$-branes at orbifold singularities.  Explicit tree-level and one-loop calculations for the $\ZZ_2$ and $\ZZ_3$ model confirm 
our expectations. It would be interesting to see if this result extends also to other types of singularities and to other types of branes.

% This expectation is supported by the fact that the gauge theory has a UV conformal fixed point. By AdS/CFT, the dual statement is that the 5-dimensional geometry (four of the dimensions along the brane, plus the radial distance) is asymptotically (at large distances) AdS$_5$.  Hence, the field theory calculation of the one and two-loop beta-functions confirms the results of a holographic analysis: asymptotically-AdS solutions are dual to theories with UV conformal fixed points, and if one turns on the fermion masses, the sum of the squares of the boson masses is  automatically determined to be equal to the sum of the squares of the fermion masses. 
%, because of the backreaction of the three-forms dual to these fermion masses. Conversely, in perturbative field theory one can turn on arbitrary boson and fermion masses, but for a generic choice of masses the beta-functions for the masses will be nonzero and the theory will not have a UV conformal fixed point. These beta-functions only vanish when the sums of the squares of the fermion and boson masses are equal. We can graphically summarize this as two equivalent statements
%\bea
%{\rm HOLOGRAPHY:} \ {\rm UV~conformal}~~~ & \rightarrow &~~~ \sum_{bosons}m_b^2 = \sum_{fermions}m_f^2 ~~~~~\nonumber \\
%{\rm FIELD~THEORY:}\sum_{bosons}m_b^2 \neq \sum_{fermions}m_f^2 ~~~ & \rightarrow & ~~~{\rm Not~UV~conformal}~~~~~ \nonumber
%\eea

\vspace*{-.4cm}
\acknowledgments{\vspace*{-.4cm}We would like to thank Stefano Massai for collaboration in the early stages of this project, and Riccardo Argurio, Johan Bl\aa b\" ack, Michele Cicoli, Joe Conlon, Csaba Cs\'aki, Emilian Dudas, Marc Goodsell, Gregory Korchemsky, Luis Iba\~nez, Carlos Savoy, Marika Taylor, David Turton, Nick Warner, Alberto Zaffaroni and Fabio Zwirner for insightful discussions. This work was supported in part by the ERC Starting Grants 240210
{\em String-QCD-BH} and 259133  {\em ObservableString} by the John Templeton Foundation Grant 48222, by the FQXi grant FQXi-RFP3-1321 (administered by Theiss Research) and by the P2IO LabEx (ANR-10-LABX-0038) in the framework Investissements d'Avenir (ANR-11-IDEX-0003-01) managed by the French ANR.}
\vspace*{.3cm}

\appendix

\section{'t Hooft symbols}\vspace*{-.4cm}
\label{sec:t-Hooft-matrices}
\noindent We have used a basis where the 't Hooft matrices $\eta^A_{ij}$ are
\begin{eqnarray} \label{GA}
\eta^1 = -i \left(\!\!
\begin{array}{cc}
0 & \sigma_2 \\
 \sigma_2 & 0
\end{array}
\!\!\right)
&
\!\!\!\!\eta^2 = \left(\!\!
\begin{array}{cc}
0 & - \sigma_0 \\
 \sigma_0 & 0
\end{array}
\!\!\right)
&
\!\!\!\!\eta^3 = i \left(\!\!
\begin{array}{cc}
 \sigma_2 & 0 \\
 0 & - \sigma_2
\end{array}
\!\!\right)
\nonumber \\
\eta^4 = i \left(\!\!
\begin{array}{cc}
0 & -  \sigma_1 \\
 \sigma_1 & 0
\end{array}
\!\!\right)
&
\eta^5 = i \left(\!\!
\begin{array}{cc}
0 &  \sigma_3 \\
 -  \sigma_3 & 0
\end{array}
\!\!\right)
&
\eta^6 = \left(\!\!
\begin{array}{cc}
\sigma_2 & 0 \\
 0 & \sigma_2
\end{array}
\!\!\right)\,,\nonumber
\end{eqnarray}
where $\sigma_{1,2,3}$ are the Pauli matrices and $\sigma_0$ is the $2\times 2$ unit matrix and we have chosen the complex coordinates
\beq \label{complexcoord}
z^i=\frac{1}{\sqrt 2}(x^1+i x^4, x^2+ix^5,x^3+ix^6) \ .\nonumber
\eeq

%The matrices $\eta^A$ satisfy the following basis-independent properties:
%\begin{equation*}
%\label{eq:G-properties-1}
%\begin{split}
%& \delta_{A B} \eta^A_{ij}   \eta^{B \, kl} = -2 \left( \delta_i^k \delta_j^l - \delta_j^k \delta_i^l  \right) \\
%&\textrm{Tr} \left( {\eta^A}^\dagger \eta^B \right) = \eta^{A \, ij} \eta^{B \, ji}  =  4 \delta_{A B} \, \\
%&\eta_{ik}^A (\eta^{B \, k j})^\dagger  + \eta_{ik}^B  (\eta^{A \, k j})^\dagger  = 2 \delta^{A B} \delta^j_i \\
%&i \epsilon_{ABCDEF}{\eta^A}_{i k_1}  \eta^{B \, k_1 k_2}  \eta^{C}_{k_2 k_3}  \eta^{D \, k_4 k_5} \eta^E_{k_5 k_6}  \eta^{F \, k_6 j}  = \delta^j_i 
%\end{split}
%\end{equation*}
%

%\vspace*{-.8cm}

\bibliography{Draft}

\begin{thebibliography}{22}
\expandafter\ifx\csname natexlab\endcsname\relax\def\natexlab#1{#1}\fi
\expandafter\ifx\csname bibnamefont\endcsname\relax
  \def\bibnamefont#1{#1}\fi
\expandafter\ifx\csname bibfnamefont\endcsname\relax
  \def\bibfnamefont#1{#1}\fi
\expandafter\ifx\csname citenamefont\endcsname\relax
  \def\citenamefont#1{#1}\fi
\expandafter\ifx\csname url\endcsname\relax
  \def\url#1{\texttt{#1}}\fi
\expandafter\ifx\csname urlprefix\endcsname\relax\def\urlprefix{URL }\fi
\providecommand{\bibinfo}[2]{#2}
\providecommand{\eprint}[2][]{\url{#2}}

\bibitem[{\citenamefont{Aldazabal et~al.}(2000)\citenamefont{Aldazabal, Ibanez,
  Quevedo, and Uranga}}]{braneSM1}
\bibinfo{author}{\bibfnamefont{G.}~\bibnamefont{Aldazabal}},
  \bibinfo{author}{\bibfnamefont{L.~E.} \bibnamefont{Ibanez}},
  \bibinfo{author}{\bibfnamefont{F.}~\bibnamefont{Quevedo}}, \bibnamefont{and}
  \bibinfo{author}{\bibfnamefont{A.~M.} \bibnamefont{Uranga}},
  \bibinfo{journal}{JHEP} \textbf{\bibinfo{volume}{08}}, \bibinfo{pages}{002}
  (\bibinfo{year}{2000}), \eprint{hep-th/0005067}.

\bibitem[{\citenamefont{Blumenhagen et~al.}(2007)\citenamefont{Blumenhagen,
  Kors, Lust, and Stieberger}}]{braneSM2}
\bibinfo{author}{\bibfnamefont{R.}~\bibnamefont{Blumenhagen}},
  \bibinfo{author}{\bibfnamefont{B.}~\bibnamefont{Kors}},
  \bibinfo{author}{\bibfnamefont{D.}~\bibnamefont{Lust}}, \bibnamefont{and}
  \bibinfo{author}{\bibfnamefont{S.}~\bibnamefont{Stieberger}},
  \bibinfo{journal}{Phys. Rept.} \textbf{\bibinfo{volume}{445}},
  \bibinfo{pages}{1} (\bibinfo{year}{2007}), \eprint{hep-th/0610327}.

\bibitem[{\citenamefont{Conlon et~al.}(2009)\citenamefont{Conlon, Maharana, and
  Quevedo}}]{braneSM3}
\bibinfo{author}{\bibfnamefont{J.~P.} \bibnamefont{Conlon}},
  \bibinfo{author}{\bibfnamefont{A.}~\bibnamefont{Maharana}}, \bibnamefont{and}
  \bibinfo{author}{\bibfnamefont{F.}~\bibnamefont{Quevedo}},
  \bibinfo{journal}{JHEP} \textbf{\bibinfo{volume}{05}}, \bibinfo{pages}{109}
  (\bibinfo{year}{2009}), \eprint{0810.5660}.

\bibitem[{\citenamefont{Gra\~na et~al.}(2004)\citenamefont{Gra\~na, Grimm,
  Jockers, and Louis}}]{Grana:2003ek}
\bibinfo{author}{\bibfnamefont{M.}~\bibnamefont{Gra\~na}},
  \bibinfo{author}{\bibfnamefont{T.~W.} \bibnamefont{Grimm}},
  \bibinfo{author}{\bibfnamefont{H.}~\bibnamefont{Jockers}}, \bibnamefont{and}
  \bibinfo{author}{\bibfnamefont{J.}~\bibnamefont{Louis}},
  \bibinfo{journal}{Nucl. Phys.} \textbf{\bibinfo{volume}{B690}},
  \bibinfo{pages}{21} (\bibinfo{year}{2004}), \eprint{hep-th/0312232}.

\bibitem[{\citenamefont{Camara et~al.}(2004)\citenamefont{Camara, Ibanez, and
  Uranga}}]{Camara:2003ku}
\bibinfo{author}{\bibfnamefont{P.~G.} \bibnamefont{Camara}},
  \bibinfo{author}{\bibfnamefont{L.~E.} \bibnamefont{Ibanez}},
  \bibnamefont{and} \bibinfo{author}{\bibfnamefont{A.~M.}
  \bibnamefont{Uranga}}, \bibinfo{journal}{Nucl. Phys.}
  \textbf{\bibinfo{volume}{B689}}, \bibinfo{pages}{195} (\bibinfo{year}{2004}),
  \eprint{hep-th/0311241}.

\bibitem[{\citenamefont{Douglas and Moore}(1996)}]{Douglas:1996sw}
\bibinfo{author}{\bibfnamefont{M.~R.} \bibnamefont{Douglas}} \bibnamefont{and}
  \bibinfo{author}{\bibfnamefont{G.~W.} \bibnamefont{Moore}}
  (\bibinfo{year}{1996}), \eprint{hep-th/9603167}.

\bibitem[{\citenamefont{Gra\~na}(2002)}]{Grana:2002tu}
\bibinfo{author}{\bibfnamefont{M.}~\bibnamefont{Gra\~na}},
  \bibinfo{journal}{Phys. Rev.} \textbf{\bibinfo{volume}{D66}},
  \bibinfo{pages}{045014} (\bibinfo{year}{2002}), \eprint{hep-th/0202118}.

\bibitem[{\citenamefont{Polchinski and Strassler}(2000)}]{Polchinski:2000uf}
\bibinfo{author}{\bibfnamefont{J.}~\bibnamefont{Polchinski}} \bibnamefont{and}
  \bibinfo{author}{\bibfnamefont{M.~J.} \bibnamefont{Strassler}}
  (\bibinfo{year}{2000}), \eprint{hep-th/0003136}.

\bibitem[{\citenamefont{Gra\~na and Polchinski}(2001)}]{Grana:2000jj}
\bibinfo{author}{\bibfnamefont{M.}~\bibnamefont{Gra\~na}} \bibnamefont{and}
  \bibinfo{author}{\bibfnamefont{J.}~\bibnamefont{Polchinski}},
  \bibinfo{journal}{Phys. Rev.} \textbf{\bibinfo{volume}{D63}},
  \bibinfo{pages}{026001} (\bibinfo{year}{2001}), \eprint{hep-th/0009211}.

\bibitem[{\citenamefont{Myers}(1999)}]{Myers:1999ps}
\bibinfo{author}{\bibfnamefont{R.~C.} \bibnamefont{Myers}},
  \bibinfo{journal}{JHEP} \textbf{\bibinfo{volume}{9912}}, \bibinfo{pages}{022}
  (\bibinfo{year}{1999}), \eprint{hep-th/9910053}.

\bibitem[{\citenamefont{Giddings et~al.}(2002)\citenamefont{Giddings, Kachru,
  and Polchinski}}]{Giddings:2001yu}
\bibinfo{author}{\bibfnamefont{S.~B.} \bibnamefont{Giddings}},
  \bibinfo{author}{\bibfnamefont{S.}~\bibnamefont{Kachru}}, \bibnamefont{and}
  \bibinfo{author}{\bibfnamefont{J.}~\bibnamefont{Polchinski}},
  \bibinfo{journal}{Phys.Rev.} \textbf{\bibinfo{volume}{D66}},
  \bibinfo{pages}{106006} (\bibinfo{year}{2002}), \eprint{hep-th/0105097}.

\bibitem[{\citenamefont{Kachru et~al.}(2003)\citenamefont{Kachru, Kallosh,
  Linde, and Trivedi}}]{KKLT}
\bibinfo{author}{\bibfnamefont{S.}~\bibnamefont{Kachru}},
  \bibinfo{author}{\bibfnamefont{R.}~\bibnamefont{Kallosh}},
  \bibinfo{author}{\bibfnamefont{A.~D.} \bibnamefont{Linde}}, \bibnamefont{and}
  \bibinfo{author}{\bibfnamefont{S.~P.} \bibnamefont{Trivedi}},
  \bibinfo{journal}{Phys. Rev.} \textbf{\bibinfo{volume}{D68}},
  \bibinfo{pages}{046005} (\bibinfo{year}{2003}), \eprint{hep-th/0301240}.

\bibitem[{\citenamefont{Conlon et~al.}(2005)\citenamefont{Conlon, Quevedo, and
  Suruliz}}]{LSV}
\bibinfo{author}{\bibfnamefont{J.~P.} \bibnamefont{Conlon}},
  \bibinfo{author}{\bibfnamefont{F.}~\bibnamefont{Quevedo}}, \bibnamefont{and}
  \bibinfo{author}{\bibfnamefont{K.}~\bibnamefont{Suruliz}},
  \bibinfo{journal}{JHEP} \textbf{\bibinfo{volume}{08}}, \bibinfo{pages}{007}
  (\bibinfo{year}{2005}), \eprint{hep-th/0505076}.

\bibitem[{\citenamefont{Kaplunovsky and Louis}(1993)}]{KL}
\bibinfo{author}{\bibfnamefont{V.~S.} \bibnamefont{Kaplunovsky}}
  \bibnamefont{and} \bibinfo{author}{\bibfnamefont{J.}~\bibnamefont{Louis}},
  \bibinfo{journal}{Phys. Lett.} \textbf{\bibinfo{volume}{B306}},
  \bibinfo{pages}{269} (\bibinfo{year}{1993}), \eprint{hep-th/9303040}.

\bibitem[{\citenamefont{Brignole et~al.}(2010)\citenamefont{Brignole, Ibanez,
  and Munoz}}]{BIM}
\bibinfo{author}{\bibfnamefont{A.}~\bibnamefont{Brignole}},
  \bibinfo{author}{\bibfnamefont{L.~E.} \bibnamefont{Ibanez}},
  \bibnamefont{and} \bibinfo{author}{\bibfnamefont{C.}~\bibnamefont{Munoz}},
  \bibinfo{journal}{Adv. Ser. Direct. High Energy Phys.}
  \textbf{\bibinfo{volume}{21}}, \bibinfo{pages}{244} (\bibinfo{year}{2010}),
  \eprint{hep-ph/9707209}.

\bibitem[{\citenamefont{Jack and Jones}(1999)}]{Jack:1999ud}
\bibinfo{author}{\bibfnamefont{I.}~\bibnamefont{Jack}} \bibnamefont{and}
  \bibinfo{author}{\bibfnamefont{D.}~\bibnamefont{Jones}},
  \bibinfo{journal}{Phys.Lett.} \textbf{\bibinfo{volume}{B457}},
  \bibinfo{pages}{101} (\bibinfo{year}{1999}), \eprint{hep-ph/9903365}.

\bibitem[{\citenamefont{Martin and Vaughn}(1994)}]{MV}
\bibinfo{author}{\bibfnamefont{S.~P.} \bibnamefont{Martin}} \bibnamefont{and}
  \bibinfo{author}{\bibfnamefont{M.~T.} \bibnamefont{Vaughn}},
  \bibinfo{journal}{Phys. Rev.} \textbf{\bibinfo{volume}{D50}},
  \bibinfo{pages}{2282} (\bibinfo{year}{1994}), \bibinfo{note}{[Erratum: Phys.
  Rev.D78,039903(2008)]}, \eprint{hep-ph/9311340}.

\bibitem[{\citenamefont{Jack and Jones}(1994)}]{JJ2}
\bibinfo{author}{\bibfnamefont{I.}~\bibnamefont{Jack}} \bibnamefont{and}
  \bibinfo{author}{\bibfnamefont{D.~R.~T.} \bibnamefont{Jones}},
  \bibinfo{journal}{Phys. Lett.} \textbf{\bibinfo{volume}{B333}},
  \bibinfo{pages}{372} (\bibinfo{year}{1994}), \eprint{hep-ph/9405233}.

\bibitem[{\citenamefont{Bena et~al.}(2016)\citenamefont{Bena, Blåbäck, and
  Turton}}]{Bena:2016fqp}
\bibinfo{author}{\bibfnamefont{I.}~\bibnamefont{Bena}},
  \bibinfo{author}{\bibfnamefont{J.}~\bibnamefont{Blåbäck}},
  \bibnamefont{and} \bibinfo{author}{\bibfnamefont{D.}~\bibnamefont{Turton}}
  (\bibinfo{year}{2016}), \eprint{1602.05959}.

\bibitem[{\citenamefont{Parkes and West}(1983)}]{Parkes:1983ib}
\bibinfo{author}{\bibfnamefont{A.}~\bibnamefont{Parkes}} \bibnamefont{and}
  \bibinfo{author}{\bibfnamefont{P.~C.} \bibnamefont{West}},
  \bibinfo{journal}{Nucl.Phys.} \textbf{\bibinfo{volume}{B222}},
  \bibinfo{pages}{269} (\bibinfo{year}{1983}).

\bibitem[{\citenamefont{Becker et~al.}(2002)\citenamefont{Becker, Becker,
  Haack, and Louis}}]{BBHL}
\bibinfo{author}{\bibfnamefont{K.}~\bibnamefont{Becker}},
  \bibinfo{author}{\bibfnamefont{M.}~\bibnamefont{Becker}},
  \bibinfo{author}{\bibfnamefont{M.}~\bibnamefont{Haack}}, \bibnamefont{and}
  \bibinfo{author}{\bibfnamefont{J.}~\bibnamefont{Louis}},
  \bibinfo{journal}{JHEP} \textbf{\bibinfo{volume}{06}}, \bibinfo{pages}{060}
  (\bibinfo{year}{2002}), \eprint{hep-th/0204254}.

\bibitem[{\citenamefont{Aparicio et~al.}(2014)\citenamefont{Aparicio, Cicoli,
  Krippendorf, Maharana, Muia, and Quevedo}}]{CQ}
\bibinfo{author}{\bibfnamefont{L.}~\bibnamefont{Aparicio}},
  \bibinfo{author}{\bibfnamefont{M.}~\bibnamefont{Cicoli}},
  \bibinfo{author}{\bibfnamefont{S.}~\bibnamefont{Krippendorf}},
  \bibinfo{author}{\bibfnamefont{A.}~\bibnamefont{Maharana}},
  \bibinfo{author}{\bibfnamefont{F.}~\bibnamefont{Muia}}, \bibnamefont{and}
  \bibinfo{author}{\bibfnamefont{F.}~\bibnamefont{Quevedo}},
  \bibinfo{journal}{JHEP} \textbf{\bibinfo{volume}{11}}, \bibinfo{pages}{071}
  (\bibinfo{year}{2014}), \eprint{1409.1931}.

\end{thebibliography}

\end{document}